\documentclass[prl,aps]{revtex4}
\begin{document}


\vskip 4mm

\centerline{\large \bf On the momentum-space approach to calculation of
one-electron}
\centerline{\large \bf energy spectra and wave functions of atomic clusters}

\vskip 2mm

\centerline{V. F. Elesin, A. I. Podlivaev, L. A. Openov}

\vskip 2mm

\centerline{\it Moscow Engineering Physics Institute
(State University),}
\centerline{\it Kashirskoe sh. 31, Moscow 115409, Russia}

\vskip 4mm

\begin{quotation}

Momentum-space approach to calculation of one-electron energies and wave
functions proposed initially by Fock for a hydrogen atom and considered later
by Shibuya, Wulfman, and Koga for diatomic molecules is applied to clusters
composed of three and more atoms. The corresponding basis set in the
coordinate space is of the Sturmian type since all the hydrogenlike orbitals
in this set have a common exponent, i.e., correspond to the same energy (as
opposed to one-electron atomic orbitals). By the examples of He$_4^{+7}$ and
He$_6^{+11}$ cluster ions it is shown that increase in the number of orbitals
in the set results in rapid convergence of eigenenergies and eigenfunctions
of highly excited states. The momentum-space approach to the one-electron
many-center problem may be used for various solid-state and quantum-chemical
applications.

\end{quotation}

\vskip 6mm

\centerline{\bf I. INTRODUCTION}

\vskip 2mm

In general, the one-electron many-center problem (i.e., the solution of the
Schr\"odinger equation for an electron moving in the field of an arbitrary
number of three-dimensional Coulomb centers) is known to have no exact
analytical solutions. Making use of the variational approach does not allow
for calculation of characteristics of excited states with controlled
accuracy. Meanwhile, the knowledge of one-particle energy spectra and wave
functions could facilitate the solution of numerous solid-state and
quantum-chemical problems.

In the works \cite{Shibuya,Koga,Koga1,Avery}, the nonvariational approach
based on the Fock representation \cite{Fock} of the momentum-space
Schr\"odinger equation has been proposed as an alternative to the usual
position-space methods. However, in the papers
\cite{Shibuya,Koga,Koga1,Avery} the momentum-space approach has been applied
to the diatomic molecules only (two-center problem).

The purpose of the present paper is to check the applicability of the
momentum-space approach to the one-electron many-center problem, i.e., to
calculation of one-electron energies and wave functions (including those for
highly excited states) of systems composed of three or more Coulomb centers.
Among other things this is important for calculations on atomic clusters. It
is worthwhile to note that the many-center problem is basically different
from the two-center one in that there appear three-center overlap integrals
which are absent in diatomics., and it is not obvious in advance if the
momentum-space approach will remain powerful in this case.

Below we present the results of calculations of the one-electron energy
spectra and wave functions of He$_4^{+7}$ and He$_6^{+11}$ cluster ions
within the momentum-space approach. The results obtained are verified by the
direct numerical solution of the coordinate-space Schr\"odinger
equation on a grid. We use atomic units $e=m=\hbar =1$ throughout the paper.

\vskip 6mm

\centerline{\bf II. MATHEMATICAL FRAMEWORK}

\vskip 2mm

In a momentum space, the Schr\"odinger equation
$[-\Delta/2-E+V({\bf r})]\Psi ({\bf r})=0$ for the energy $E$ and the wave
function $\Psi ({\bf r})$ of an electron moving in an external potential
$V({\bf r})$ is an integral equation for a Fourier transform
$\Psi ({\bf p})=(2\pi)^{-3/2}\int d{\bf r}\Psi ({\bf r})\exp(-i{\bf pr})$
of $\Psi ({\bf r})$ \cite{Shibuya,Koga,Koga1,Avery,Fock}:
\begin{equation}
(|{\bf p}|^2+|{\bf p}_0|^2)\Psi ({\bf p})=-2(2\pi)^{-3/2}
\int d{\bf p}^{\prime}V({\bf p}-{\bf p}^{\prime})\Psi ({\bf p}^{\prime})~,
\label{Schrodinger}
\end{equation}
where $|{\bf p}_0|^2=2|E|$ and
$V({\bf p})=(2\pi)^{-3/2}\int d{\bf r}V({\bf r})\exp(-i{\bf pr})$. For a
cluster composed of $N_{ion}$ ions with charges $Z_k$ and coordinates
${\bf R}_k$, where $k=1,...,N_{ion}$ is the number of an ion in the cluster,
one has $V({\bf r})=-\sum_{k}Z_k/|{\bf r}-{\bf R}_k|$ so that
$V({\bf p})=-4\pi(2\pi)^{-3/2}|{\bf p}|^{-2}\sum_{k}Z_k\exp(-i{\bf pR}_k)$,
and hence Eq. (\ref{Schrodinger}) reads
\begin{equation}
(|{\bf p}|^2+|{\bf p}_0|^2)\Psi ({\bf p})=\pi^{-2}
\sum_{k}Z_k\exp(-i{\bf pR}_k)
\int d{\bf p}^{\prime}|{\bf p}-{\bf p}^{\prime}|^{-2}
\exp(i{\bf p}^{\prime}{\bf R}_k)\Psi ({\bf p}^{\prime})~.
\label{Schrodinger2}
\end{equation}

The Fock transformation \cite{Fock} projects the three-dimensional momentum
space onto the four-dimensional sphere with a radius $|{\bf p}_0|$, where
$|{\bf p}_0|$ is the energy-related momentum that enters
Eqs. (\ref{Schrodinger}) and (\ref{Schrodinger2}). The function
$\Psi ({\bf p})$ is related to its four-dimensional image
$\Psi (\Omega)$ through
\begin{equation}
\Psi ({\bf p})=
4|{\bf p}_0|^{5/2}(|{\bf p}|^2+|{\bf p}_0|^2)^{-2}\Psi(\Omega)~,
\label{Psi_p}
\end{equation}
where $\Omega$ stands for the set of three angles $(\alpha,\theta,\phi)$
resulting from the Fock transformation of a momentum vector
${\bf p}=(|{\bf p}|,\theta,\phi)$. Here the angles $\theta$ and $\phi$ have
usual meaning and the angle $\alpha$ is related to $|{\bf p}|$ through
$\alpha=2\arctan(|{\bf p}|/|{\bf p}_0|)$ so that $0\leq\alpha<\pi$ for
$0\leq{\bf p}|<\infty$. The function $\Psi(\Omega)$ can be represented as
\begin{equation}
\Psi(\Omega)=\sum_{k,N}a_{kN}\exp(-i{\bf pR}_k)Y_N(\Omega)~,
\label{Psi_Omega}
\end{equation}
where $N=(n,l,m)$ stands for a set of three quantum numbers (principal
$n=1,2,...$, orbital $l=0,1,...,n-1$, and azimuthal $m=-l,...,l$) and
$Y_N(\Omega)$ is a four-dimensional spherical harmonic defined as
$Y_N(\Omega)=(-1)^lC_{nl}(\alpha)Y_{lm}(\theta,\phi)$. Here
$C_{nl}$ are Gegenbauer polynomials and $Y_{lm}$ are usual three-dimensional
spherical functions.

By the example of a diatomic molecule, it was shown in Ref. \cite{Koga1} that
coefficients $a_{kN}$ in Eq. (\ref{Psi_Omega}) satisfy the set of coupled
homogeneous algebraic equations. Generalizing the consideration of
Ref. \cite{Koga1} to the case of an arbitrary number $N_{ion}$ of ions in the
cluster, we have
\begin{equation}
\sum_j H_{ij}a_j=0~,
\label{a_kN}
\end{equation}
where $i=(k,N)$, $j=(k^{\prime},N^{\prime})$,
\begin{equation}
H_{ij}=|{\bf p}_0|S_N^{N^{\prime}}({\bf R}_k-{\bf R}_{k^{\prime}})-
\sum_{k^{\prime\prime}}Z_{k^{\prime\prime}}\sum_{N^{\prime\prime}}
\frac{1}{n^{\prime\prime}}
S_N^{N^{\prime\prime}}({\bf R}_k-{\bf R}_{k^{\prime\prime}})
S_{N^{\prime\prime}}^{N^{\prime}}
({\bf R}_{k^{\prime\prime}}-{\bf R}_{k^{\prime}})~,
\label{Hij}
\end{equation}
\begin{equation}
S_N^{N^{\prime}}({\bf R}_k-{\bf R}_{k^{\prime}})=
\int d\Omega Y_N^*(\Omega)Y_{N^{\prime}}(\Omega)
\exp[i{\bf p}({\bf R}_k-{\bf R}_{k^{\prime}})]~.
\label{S_Omega}
\end{equation}
The function $S_N^{N^{\prime}}({\bf R}_k-{\bf R}_{k^{\prime}})$ can be
expressed \cite{Shibuya,Koga,Koga1} as an integral
\begin{equation}
S_N^{N^{\prime}}({\bf R}_k-{\bf R}_{k^{\prime}})=
\frac{n}{|{\bf p}_0|}\int d{\bf r}\frac
{\chi_N^*({\bf r}-{\bf R}_k)\chi_{N^{\prime}}({\bf r}-{\bf R}_{k^{\prime}})}
{|{\bf r}-{\bf R}_k|}
\label{S_chi}
\end{equation}
over localized hydrogenlike orbitals
\begin{equation}
\chi_N({\bf r})=(-1)^{n-l-1}2|{\bf p}_0|^{3/2}[(n-l-1)!/n(n+l)!]^{1/2}
(2t)^l\exp(-t)L_{n-l-1}^{2l+1}(2t)Y_{lm}(\theta,\phi)~,
\label{chi}
\end{equation}
where $t=|{\bf p}_0||{\bf r}|$ and $L_i^j$ are the associated Laguerre
polynomials. Note that all orbitals $\chi_N({\bf r})$ have a common exponent
$|{\bf p}_0|$, contrary to one-electron atomic wave functions. The functions
$\chi_N({\bf r})$ form the Sturmian basis set which, as noticed in
Ref. \cite{Avery}, is a basis of a Sobolev space rather than a Hilbert space.
All orbitals $\chi_N({\bf r})$ correspond to the same energy, regardless of
their quantum numbers.

Eq. (\ref{a_kN}) is the nonvariational equation. To find the expansion
coefficients $a_{kN}$ and the energy parameter $|{\bf p}_0|$, one should
solve the nonlinear equation $\det(H_{ij})=0$ for $|{\bf p}_0|$. Different
solutions $|{\bf p}_0|_i$ of this equation correspond to different
eigenenergies $E_i$ and eigenfunctions $\Psi_i({\bf r})$ of the one-electron
Schr\"odinger equation in the coordinate space. Once the values of
$|{\bf p}_0|$ and $a_{kN}$ are obtained, the corresponding electron energy
and wave function are, respectively, $E=-|{\bf p}_0|^2/2$ and \cite{Norma}
\begin{equation}
\Psi({\bf r})=\sum_{k,N}a_{kN}\chi_N({\bf r}-{\bf R}_k)~.
\label{Psi_r}
\end{equation}

In general, there are four types of integrals over orbitals $\chi_N({\bf r})$
in the matrix elements $H_{ij}$, see Eqs. (\ref{Hij}) and (\ref{S_chi}).
First is a one-center integral
\begin{equation}
\int d{\bf r}\frac{\chi_N^*({\bf r})\chi_{N^{\prime}}({\bf r})}{|{\bf r}|}~.
\label{one-center}
\end{equation}
It appears if $k=k^{\prime}$, see Eq. (\ref{S_chi}), and equals to
$(|{\bf p}_0|/n)\delta_{NN^{{\prime}}}$, so that
$S_N^{N^{\prime}}(0)=\delta_{NN^{{\prime}}}$. Second is a two-center integral
\begin{equation}
\int d{\bf r}\frac{\chi_N^*({\bf r})\chi_{N^{\prime}}({\bf r}-{\bf R})}
{|{\bf r}|}
\label{two-center}
\end{equation}
that enters $S_N^{N^{\prime}}({\bf R}_k-{\bf R}_{k^{\prime}})$ at
$k\neq k^{\prime}$ (here ${\bf R}={\bf R}_{k^{\prime}}-{\bf R}_k$). Next, it
can be shown \cite{Koga1} that at $k=k^{\prime}\neq k^{\prime\prime}$ the sum
$\sum_{N^{\prime\prime}}$ in the second term of Eq. (\ref{Hij}) equals to
another two-center integral
\begin{equation}
\int d{\bf r}\frac{\chi_N^*({\bf r})\chi_{N^{\prime}}({\bf r})}
{|{\bf r}-{\bf R}|}
\label{two-center2}
\end{equation}
divided by $|{\bf p}_0|$, where
${\bf R}={\bf R}_{k^{\prime\prime}}-{\bf R}_k$. Finally, at
$k\neq k^{\prime}$, $k\neq k^{\prime\prime}$, and
$k^{\prime}\neq k^{\prime\prime}$, the sum
$\sum_{N^{\prime\prime}}$ in Eq. (\ref{Hij}) is a three-center integral
\begin{equation}
\int d{\bf r}\frac{\chi_N^*({\bf r})\chi_{N^{\prime}}({\bf r}-{\bf R}_1)}
{|{\bf r}-{\bf R}_2|}
\label{three-center}
\end{equation}
divided by $|{\bf p}_0|$, where ${\bf R}_1={\bf R}_{k^{\prime}}-{\bf R}_k$
and ${\bf R}_2={\bf R}_{k^{\prime\prime}}-{\bf R}_k$.

In the case of a single ion ($N_{ion}=1$) with a charge $Z$, there are only
one-center integrals (\ref{one-center}) in the matrix elements $H_{ij}$, and
Eq. (\ref{a_kN}) allows for a simple analytical solution \cite{Fock}. Since
$H_{ij}=(|{\bf p}_0|-Z/n)\delta_{NN^{\prime}}$ at $k=k^{\prime}$, one has
$|{\bf p}_0|=Z/n$ and $E=-Z^2/2n^2$.

As one can see from Eqs. (\ref{Hij}) and (\ref{S_chi}), for a diatomic
molecule ($N_{ion}=2$), the matrix elements $H_{ij}$ include both one-center
(\ref{one-center}) and two-center (\ref{two-center}), (\ref{two-center2})
integrals. The analytical expressions for two-center integrals are rather
complex \cite{Avery}. Hence, in order to solve Eq. (\ref{a_kN}) numerically,
one should restrict himself to a finite number $M$ of orbitals
$\chi_N({\bf r})$ in the basis set. The authors of Ref. \cite{Koga1} applied
the momentum-space approach to the molecular ion H$_2^+$. They calculated the
energies of the ground ($1s\sigma_g$) and one of the excited ($2p\sigma_u$)
states and showed that, as $M$ increases, both those energies decrease and
converge to the corresponding 'exact' values obtained by the analytical
series solution in the coordinate space (see, e.g., Ref. \cite{Bates}).

\vskip 6mm

\centerline{\bf III. RESULTS AND DISCUSSION}

\vskip 2mm

If the number of ions in the system is greater than two ($N_{ion}\geq 3$),
there appear three-center integrals (\ref{three-center}) in the matrix
elements $H_{ij}$. In general, there are no exact analytical expressions for
three-center integrals. However, because of the long-range nature of the
Coulomb interaction, they are of the order of the corresponding two-center
integrals. Hence, they cannot be neglected and have to be accounted for on an
equal footing.

In order to check the applicability of the momentum-space approach to the
one-electron problem in the case $N_{ion}\geq 3$, we have calculated the
one-electron characteristics of several clusters with different ion charges
and configurations. Below we present the results for chains composed of
$N_{ion}$ = 4 and 6 ions with charge $Z=2$ each. From the one-electron
viewpoint, these systems correspond to cluster ions He$_4^{+7}$ and
He$_6^{+11}$ respectively. The linear form of the clusters was taken in order
to facilitate the visualization of the wave functions in the coordinate
space. We stress that our purpose was not to compute the energies and wave
functions with an extremely high accuracy, as in Ref. \cite{Koga1} for a
diatomic molecule, but just (i) to clarify the very possibility to apply the
momentum-space approach to atomic clusters, i.e., to the one-electron
many-center problem and (ii) to study the convergence rate at the initial
stage of increase in the number $M$ of orbitals $\chi_N({\bf r})$ in the basis
set. So, we have restricted ourselves to $M$ = 5, 14, and 23. These values of
$M$ correspond to account for all orbitals with $n\leq 2$; $n\leq 3$; and
$n\leq 4, l\leq 3$ respectively. For each $M$, we have calculated the
energies and the coordinate-space wave functions of $20 \div 30$ one-electron
levels.

In Table I, we list the one-electron energies for the first 15 levels of the
chain He$_4^{+7}$. One can see that the energy of each level decreases with
$M$, this decrease being more pronounced for highly excited states. As
one goes from $M=5$ to $M=14$, the energies of the lower four levels change
by $(0.3\div 0.5)$ \%, while the energies of the higher levels change by
$(9\div 17)$ \%. This is obviously due to larger weights of orbitals
$\chi_N({\bf r})$ with $n\geq 3$ in the wave functions of highly excited
states. Note, however, that further increase in $M$ up to $M=23$ results in
an order of magnitude weaker change of the overall energy spectrum, less than
by $0.015$ \% for the lower four levels and by $(0.3\div 2)$ \% for other
levels. So, the convergence rate is very high, being comparable to that
reported in Ref. \cite{Koga1} for H$_2^+$ molecule. Roughly speaking, an
increase in the principal quantum number $n$ by one results in convergence of
at least one more decimal digit, i.e., in at least an order of magnitude
increase in the accuracy.

By the direct numerical solution of the coordinate-space Schr\"odinger
equation on a grid, we have verified that all eigenenergies considered
indeed converge to their 'exact' values. This is true for the eigenfunctions
as well. Figure 1 shows the wave functions of the 5-th level calculated for
different values of $M$. One can see that the wave functions for $M=5$ and
$M=14$ differ considerably, while those for $M=14$ and $M=23$ are very close
to each other. It is noteworthy that the wave function for $M=23$ is visually
indistinguishable from the 'exact' wave function obtained by the direct
numerical integration of the Schr\"odinger equation. Note also that the small
basis set ($M=5$) is insufficient for even qualitative description of
highly excited states. As one can see from Table I, in the case $M=5$, the
order of levels in the energy spectrum appears to be broken starting with the
14-th level. The 14-th and 15-th levels are degenerate for $M=5$, the
corresponding wave functions being equal to zero at the line connecting the
ions in the chain. Meanwhile, those levels are non-degenerate for both $M=14$
and $M=23$, in accordance with the 'exact' numerical solution. Figure 2 shows
the wave functions of the 14-th level calculated for $M=14$ and $M=23$. They
differ considerably, while the wave function for $M=23$ practically coincides
with the 'exact' wave function.

Figure 3 presents the one-electron energies of four lowest levels of the
chain He$_4^{+7}$ as a function of the distance $R_{23}$ between two inner
ions, the values of $R_{12}$ and $R_{34}$ being fixed at $R_{12}=R_{34}=2.4$.
The first and the second cluster levels emerge from the first levels of the
diatomic He$_2^{+3}$. As $R_{23}$ decreases, the energy separation $E_2-E_1$
increases due to stronger hybridization between molecular orbitals. The same
is true for the splitting $E_4-E_3$ between the fourth and the third cluster
levels which both emerge from the second levels of the diatomic He$_2^{+3}$
as they approach each other. Note that $E_2-E_1$ at a given $R_{23}$ is much
greater than $E_4-E_3$. This is because of the different symmetry of the
first and second diatomic orbitals (symmetrical and antisymmetrical,
respectively). We have also solved the Schr\"odinger equation numerically at
several values of $R_{23}$, making use of the finite difference method on a
fine grid, and verified the results obtained by the momentum-space approach.

Finally, we calculated the one-electron energies and wave functions for the
first 29 levels of the chain He$_6^{+11}$. The results are similar to those
presented above for the chain He$_4^{+7}$. Increase in $M$ leads to the
progressive decrease of all eigenenergies and their rapid convergence to the
corresponding 'exact' values. Figure 4 shows the wave functions of the 10-th
level computed for different values of $M$. Again, as in the case of the
chain He$_4^{+7}$, the wave functions for $M=5$ and $M=14$ differ
considerably, while those for $M=14$ and $M=23$ are very close to each other,
the wave function for $M=23$ being almost identical with the 'exact' wave
function. The eigenenergies are $E_{10}=$ -2.414, -2.828, and -2.872 for
$M=$ 5, 14, and 23 respectively.

\vskip 6mm

\centerline{\bf IV. CONCLUSIONS}

\vskip 2mm

In summary, we made use of the nonvariational momentum-space approach to
calculate the one-electron energy spectra and wave functions of the ground
and a large number of excited states of small atomic clusters. We have found
that the one-electron characteristics converge rapidly with increase in the
number of hydrogenlike orbitals in the basis set, each orbital having the
same exponent, i.e., corresponding to the same energy. Our results show that
the momentum-space approach to the one-electron many-center problem is rather
powerful and may be considered as an interesting alternative to the
position-space methods.

\vskip 6mm

{\bf Acknowledgments}

\vskip 2mm

We are grateful to N. E. L'vov for valuable discussions. The work was
supported by the CRDF (Project "Basic studies of matter in extreme states"),
the DTRA (Contract No 01-02-P0280), and by the Russian Federal Program
"Integration" (Project No B0049).


\newpage

\vskip 2mm

Table I. The one-electron energy spectra of the chain He$_4^{+7}$ calculated
for different numbers $M$ of hydrogenlike orbitals $\chi_N({\bf r})$ in the
basis set. The value $M$ = 5, 14, and 23 corresponds to the basis set
composed of (1S,2S,2P), (1S,2S,2P,3S,3P,3D), and (1S,2S,2P,3S,3P,3D,4S,4P,4D)
orbitals respectively. The distance between two inner ions is $R_{23}=3$, the
distance between each outer ion and its neighbour is $R_{12}=R_{34}=2.4$. The
energies and lengths are measured in atomic units.

\vskip 6mm

\vbox{\offinterlineskip
\hrule
\halign{&\vrule#&
\strut\quad\hfil#\quad\quad\quad\cr
height2pt&\omit&&\omit&&\omit&&\omit&\cr
&Level number\hfil&&$M=5$&&$M=14$&&$M=23$&\cr
\noalign{\hrule}
height2pt&\omit&&\omit&&\omit&&\omit&\cr
& 1&&-3.8961&&-3.9151&&-3.9157&\cr
& 2&&-3.8460&&-3.8584&&-3.8586&\cr
& 3&&-3.4430&&-3.4543&&-3.4543&\cr
& 4&&-3.4348&&-3.4433&&-3.4434&\cr
& 5&&-2.0881&&-2.3063&&-2.3149&\cr
& 6&&-1.9496&&-2.2043&&-2.2237&\cr
& 7&&-1.9496&&-2.2043&&-2.2237&\cr
& 8&&-1.8304&&-2.0570&&-2.0678&\cr
& 9&&-1.8304&&-2.0570&&-2.0678&\cr
&10&&-1.7869&&-1.9504&&-1.9848&\cr
&11&&-1.6536&&-1.9431&&-1.9532&\cr
&12&&-1.4881&&-1.7440&&-1.7712&\cr
&13&&-1.4881&&-1.7440&&-1.7712&\cr
&14&&-1.4177&&-1.6308&&-1.6657&\cr
&15&&-1.4177&&-1.5386&&-1.5621&\cr
height2pt&\omit&&\omit&&\omit&&\omit&\cr}
\hrule}

\newpage

\centerline{\bf Figure captions}

\vskip 2mm

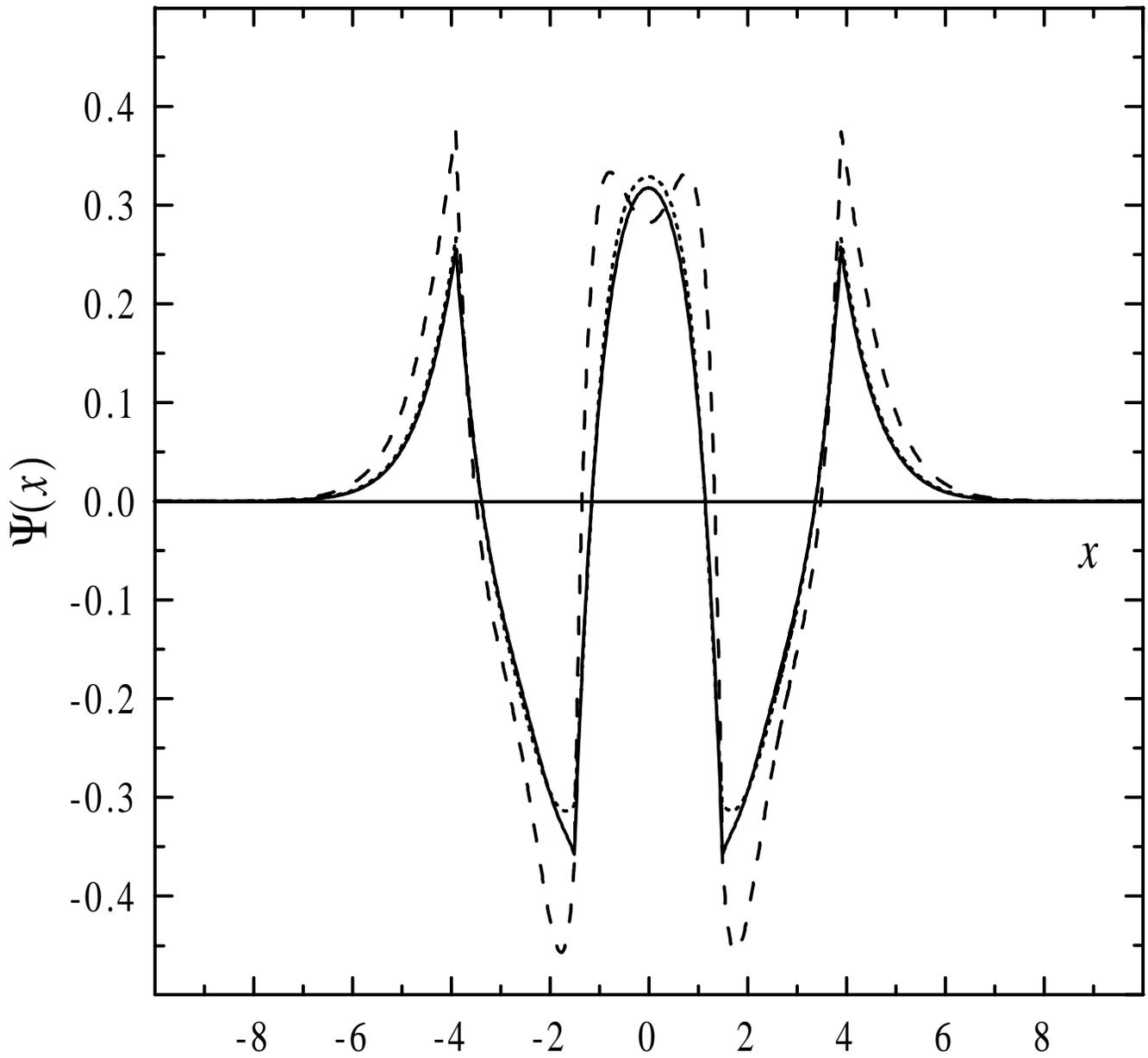
Fig. 1. The wave functions $\Psi(x)$ of the 5-th one-electron level of the
chain He$_4^{+7}$ calculated for different numbers $M$ of hydrogenlike
orbitals $\chi_N({\bf r})$ in the basis set, $M=5$ (dashed line), $M=14$
(dotted line), and $M=23$ (solid line). The ion coordinates are (-3.9, 0, 0),
(-1.5, 0, 0), (1.5, 0, 0), and (3.9, 0, 0) in atomic units. For $M=23$, the
wave function practically coincides with that obtained by the direct
numerical integration of the coordinate-space Schr\"odinger equation.

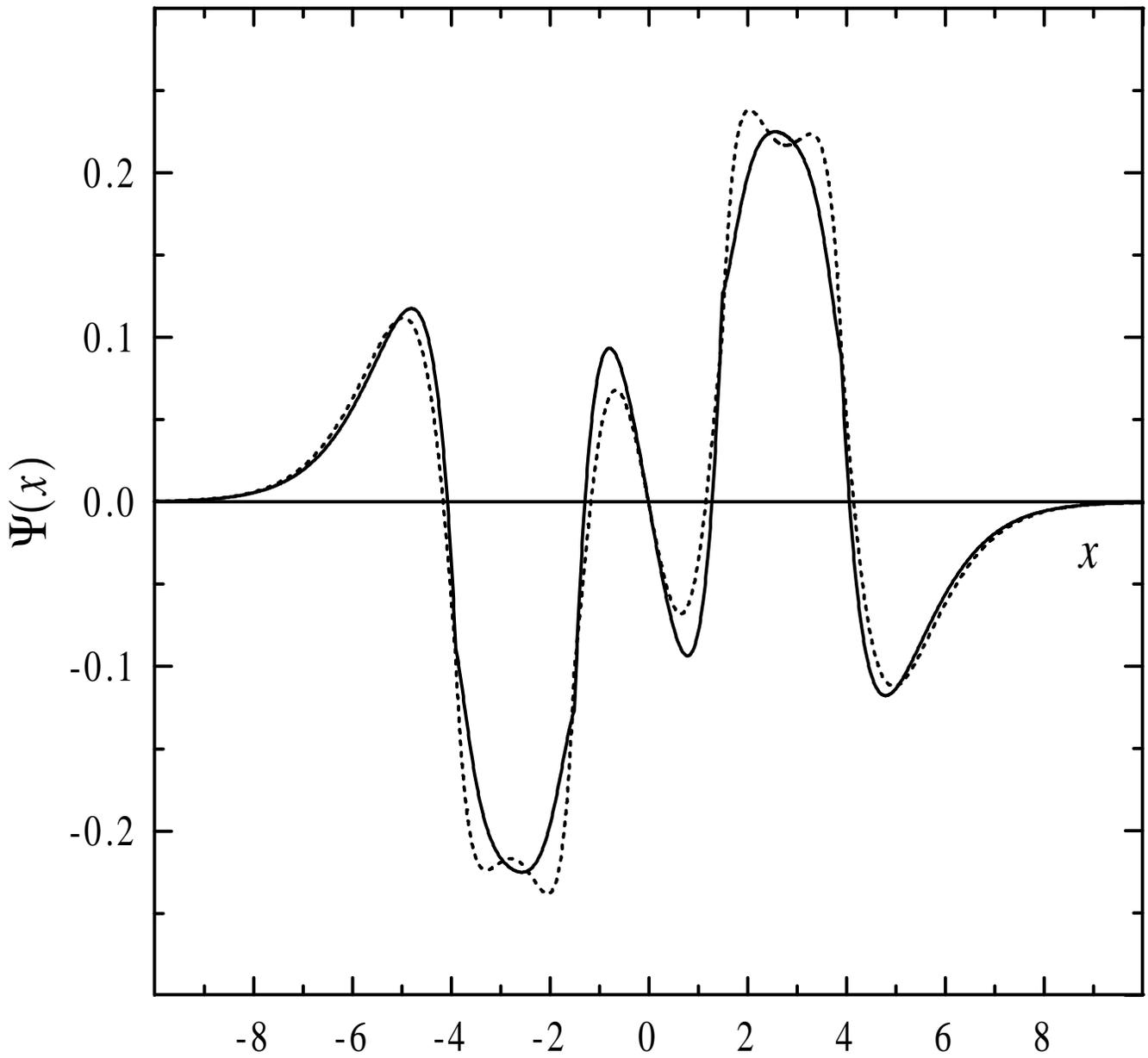
Fig. 2. The same as in Fig.1, for the 14-th one-electron level of the
chain He$_4^{+7}$. For $M=5$, the wave function is zero at $y=0$ and $z=0$,
see the text. For $M=23$, the wave function practically coincides with that
obtained by the direct numerical integration of the coordinate-space
Schr\"odinger equation.

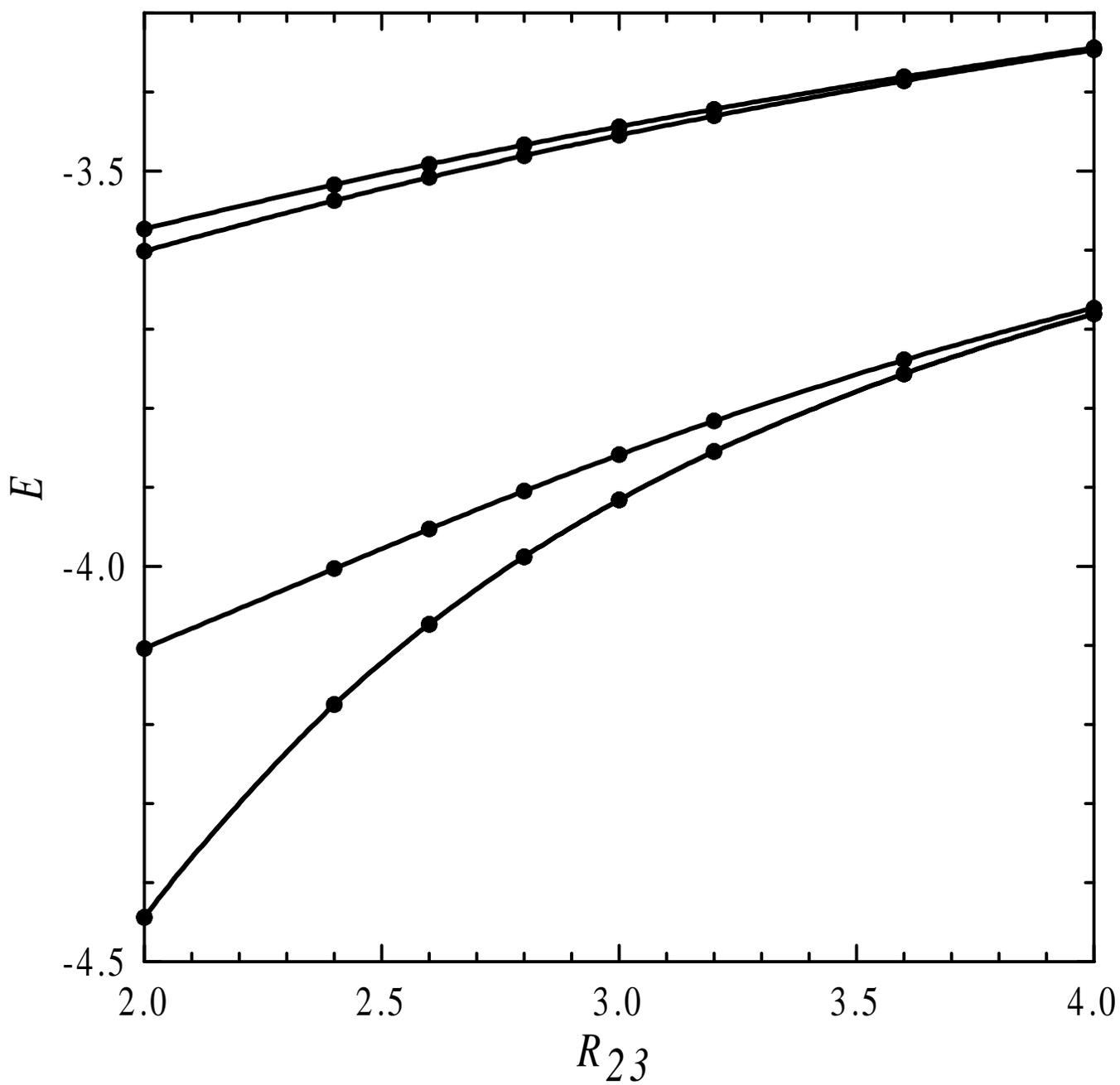
Fig. 3. The one-electron energies of the first four levels of the chain
He$_4^{+7}$ versus the distance $R_{23}$ between two inner ions, the values
of $R_{12}$ and $R_{34}$ being fixed at 2.4 each. Solid lines are the results
of the momentum-space approach for $M=23$ orbitals in the basis set. Circles
are the numerical solutions of the Schr\"odinger equation on a grid.

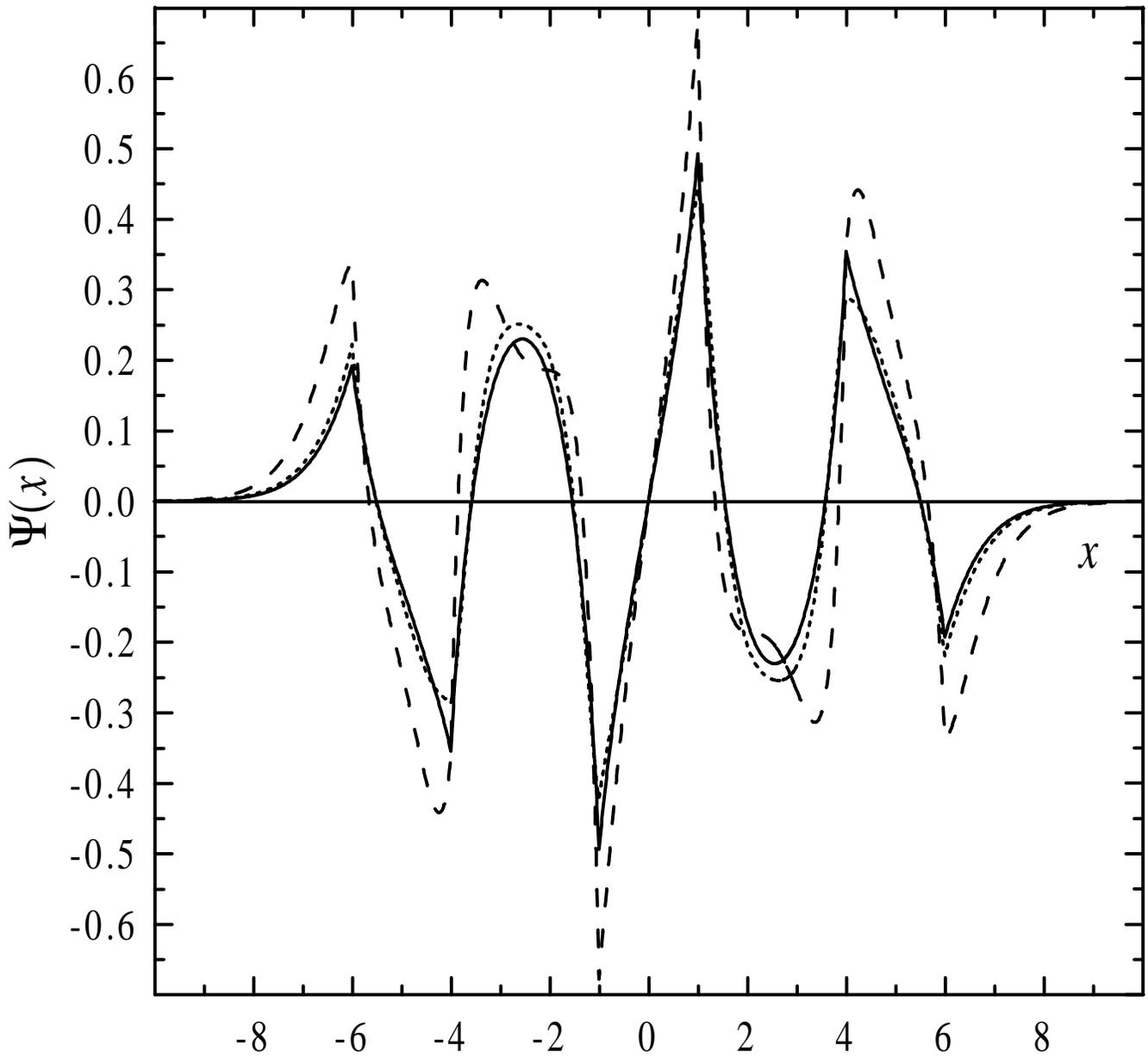
Fig. 4. The wave functions $\Psi(x)$ of the 10-th one-electron level of the
chain He$_6^{+11}$ calculated for different numbers $M$ of hydrogenlike
orbitals $\chi_N({\bf r})$ in the basis set, $M=5$ (dashed line), $M=14$
(dotted line), and $M=23$ (solid line). The ion coordinates are (-6, 0, 0),
(-4, 0, 0), (-1, 0, 0), (1, 0, 0), (4, 0, 0), and (6, 0, 0) in atomic units.
For $M=23$, the wave function practically coincides with that obtained by the
direct numerical integration of the coordinate-space Schr\"odinger equation.


\begin{references}
\bibitem{Norma} The coefficients $a_{kN}$ in the expansion of the wave
function $\Psi({\bf r})$ over the orbitals $\chi_N({\bf r})$ should be scaled
so that $\Psi({\bf r})$ was normalized to unity.
\end{references}
\end{document}